\documentclass{iau} 
\usepackage{graphicx}
\usepackage[utf8]{inputenc}
\usepackage[english]{babel}

\title[Spectroscopic properties of radio-loud and radio-quiet quasars]{Spectroscopic properties of radio-loud and radio-quiet quasars}

\author[A. Chakraborty \& A. Bhattacharjee]   
{Avinanda Chakraborty$^1$
 \and Anirban Bhattacharjee$^2$}
\affiliation{$^1$Presidency University, Kolkata, \\ Pincode 700073,
86/1 College Street, West Bengal, India \\ email: {\tt avinanda.rs@presiuniv.ac.in} \\[\affilskip]
$^2$Sul Ross State University, Texas, Box 
C-64, Alpine, TX 79832, Texas, USA \\email: {\tt anirbanbhattacharjeee@gmail.com}}

\pubyear{2019}
\volume{356}  
\jname{Nuclear Activity in Galaxies Across Cosmic Time}
\editors{M. Povi\'c, P. Marziani, J. Masegosa, H. Netzer,\\ S. H. Negu, \&
	S. B. Tessema, eds.}
\begin{document}

\maketitle
\begin{abstract}
Surveys have shown radio-loud (RL) quasars constitute 10\%-15\% of the total quasar population and rest are radio-quiet (RQ). However, it is unknown if this radio-loud fraction (RLF) remains consistent among different parameter spaces. This study shows that RLF increases for increasing full width half maximum (FWHM) velocity of the H$\beta$ broad emission line ($z <$ 0.75). To analyse the reason, we compared bolometric luminosity of RL and RQ quasars sample which have FWHM of H$\beta$ broad emission line greater than 15000km/s (High Broad Line or HBL) with which have FWHM of H$\beta$ emission line less than 2500km/s (Low Broad Line or LBL). From the distributions we can conclude for the HBL, RQ and RL quasars are peaking separately and RL quasars are having higher values whereas for the LBL the peaks are almost indistinguishable. We predicted selection effects could be the possible reason but to conclude anything more analysis is needed. Then we compared our result with Wills $\&$ Browne (1986) and have shown that some objects from our sample do not follow the pattern of the logR vs FWHM plot where R is the ratio of 5 GHz radio core flux density with the extended radio lobe flux density.  
\end{abstract}

\firstsection 
\section{Introduction}
Quasars are the most luminous active galactic nuclei (AGN) and are powered by accretion of supermassive black holes (SMBHs) (Salpeter 1964; Lynden-Bell 1969). We still don't understand properly why some active galactic nuclei have strong radio sources and others do not (Lynden-Bell 1969). Surveys have shown that other than radio surveys there is no such difference between RLQs and RQQs (Kratzer $\&$ Richards 2015). Although Radio-loud quasars (RLQs) were first detected as radio sources. Only 10\% of the total quasars are RL (Sandage 1965). The main difference between both RLQs and Radio-quiet quasars (RQQs) is the presence of powerful radio jets (e.g. Bridle et al. 1994; Mullin et al. 2008). However, there is evidence of weak radio jets in RQQs also (Ulvestad et al. 2005; Leipski et al. 2006). 
\\Wills $\&$ Browne (1986) found a significant correlation between the full width half maximum (FWHM) and of broad H$\beta$ lines and the logR where R is defined as ``{Ratio of 5 GHz core to extended component flux density} by Wills $\&$ Browne (1986).'' The parameter R has been used as a measure of orientation. These authors have shown the distribution of logR to be highly asymmetric and biased toward small H$\beta$ FWHMs with a cut off near 2000km/s.
\\There is evidence that the broad-line width measurement in quasar is dependant on the source orientation and consistent with the idea of flattened or disc like broad-line regions (Jarvis $\&$ McLure 2006). These authors have also presented a significant correlation between radio spectral index and broad-line width of the H$\beta$ and Mg II emission lines ($\gg$ 99.99$\%$). These authors showed spectral index can be used as a proxy for source orientation.
\\It has also been shown that normalizing the radio core luminosity by the optical continuum luminosity (log$R_{\mathrm{v}}$) (K-corrected) is a superior orientation indicator (Van Gorkom et al. 2015). Van Gorkom et al. 2015 compared between logR and log$R_{\mathrm{v}}$  and two other indicators of orientation, the ratio of the optical continuum luminosity and emission-line luminosity (Yee $\&$ Oke (1978)) and the ratio of the jet power and the luminosity of the narrow-line region Rawlings $\&$ Saunders (1981)).
\begin{figure}
\begin{center}
\includegraphics[width=3.5in]{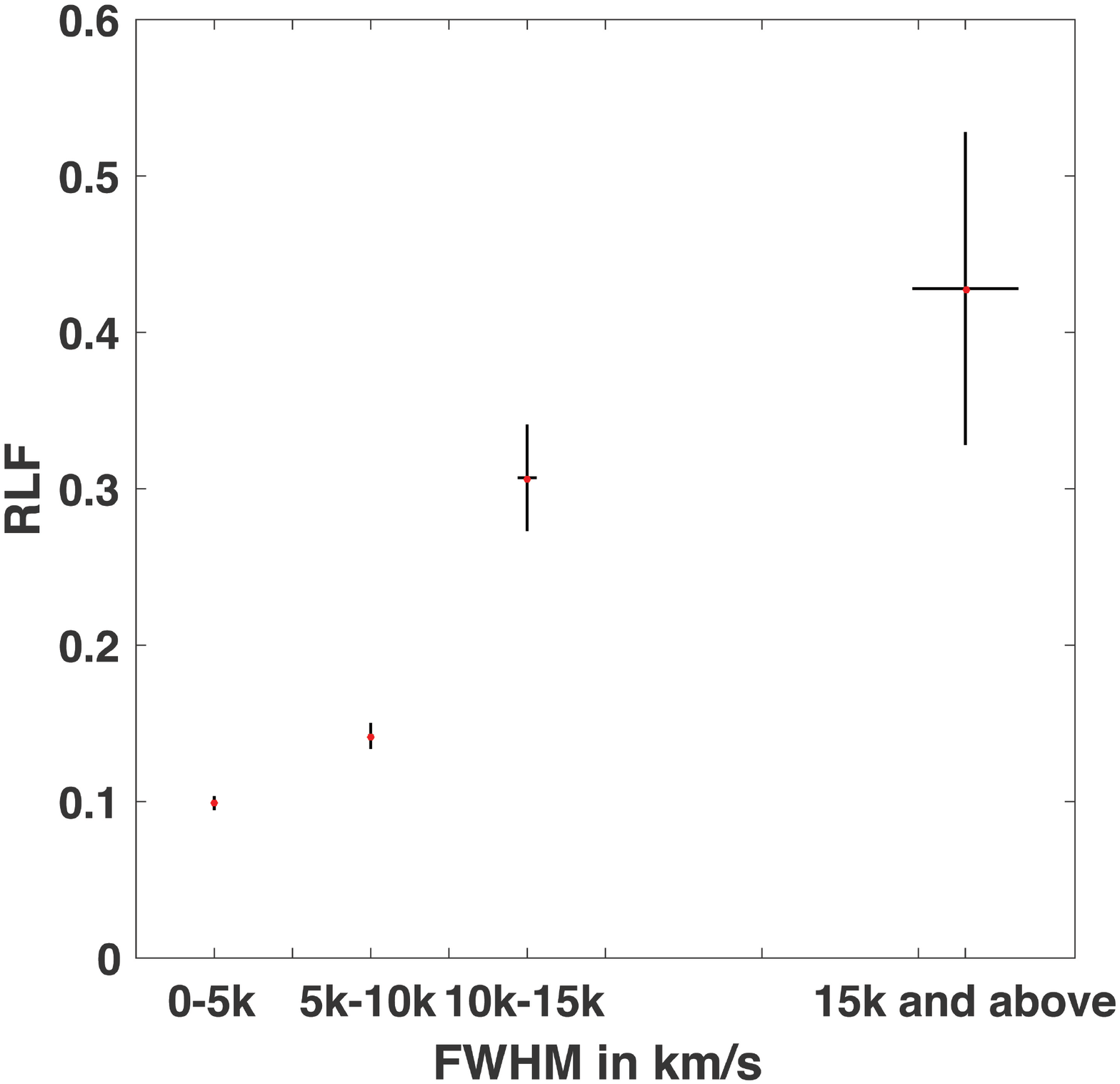}
\caption{Variation of Radio-Loud Fraction across different FWHM of broad H$\beta$ with FWHM. From this plot we can see that the RLF increases with FWHM.}
\label{Fig.1}
\end{center}
\end{figure}
\section{Overview} 
Our work is basically focused on investigating different properties of RLQs and RQQs to find the reason behind the high value of RLF at higher FWHM and comparing our results with some other literature and reach to some conclusion. Initial work was done by Bhattacharjee, Gilbert, $\&$ Brotherton (2018). To check the consistency in different parameter spaces they first looked for the variation of RLF with FWHM of broad H$\beta$. We then analysed fundamental H$\beta$ line properties of RLQs and RQQs for the HBL region (FWHM $>$ 15000km/s) and compared them with the LBL (FWHM $<$ 2500km/s)region properties to check the reason of high RLF for higher FWHM and lastly we compared our result with other literature. Here are the data samples we have used :
\\{\underline{\it Sloan Digital Sky Survey (SDSS):}}
Our main quasar catalogue comes from the SDSS (York et al. 2000) Data Release 7 (Abazajian et al. 2009) Quasar catalogue (Shen et al. 2011). It consists of 105,783 quasars brighter than $M_{i}$ = -22.0 and are spectroscopically confirmed.
\\{\underline{\it Faint Images of the Radio Sky at Twenty-cm (FIRST):}}
Shen et al. 2011 cross matched the quasar catalogue of SDSS DR7 and FIRST survey of VLA. The quasars having only one FIRST source within 5” are classified as core-dominated radio sources and those having multiple FIRST sources within 30” are classified as lobe dominated. These two categories are together named as RLQs by Shen et al. 2011. And those with only one FIRST match between 5" and 30" are classified as RQQs.  We checked optical spectra of the HBL and LBL quasars from SDSS and quasars with some issues with their H$\beta$ line are manually discarded from our sample. So our final H$\beta$ sample contains 298 RLQs and 1,910 RQQs. Among RLQs 56 are HBL (FWHM $>$ 15000 km/s) and 242 are LBL (FWHM $<$ 2500 km/s) and in RQQs 41 are HBL and 1869 are LBL sources.
\begin{figure}[tb]
\includegraphics[width=2.75in]{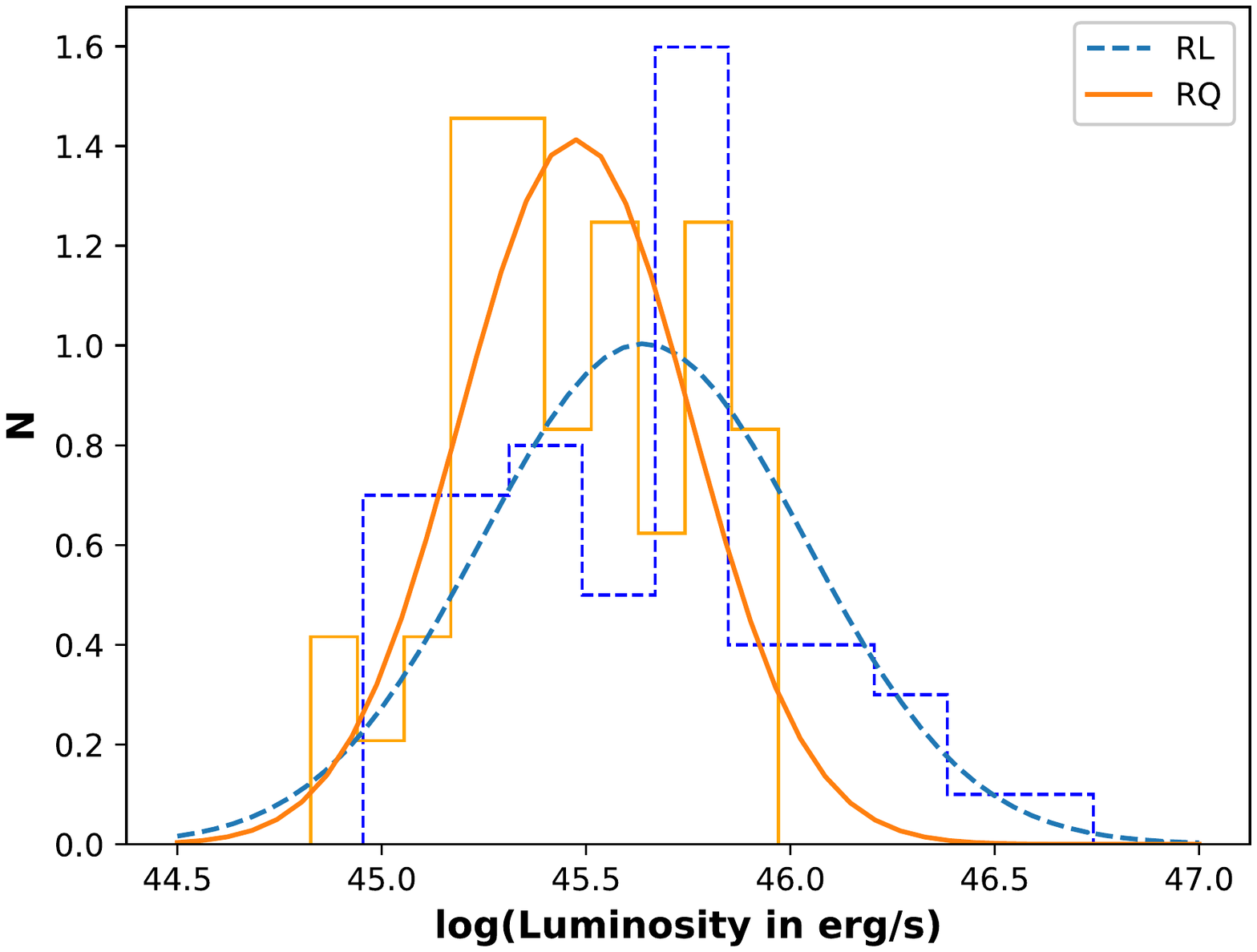}
\includegraphics[width=2.75in]{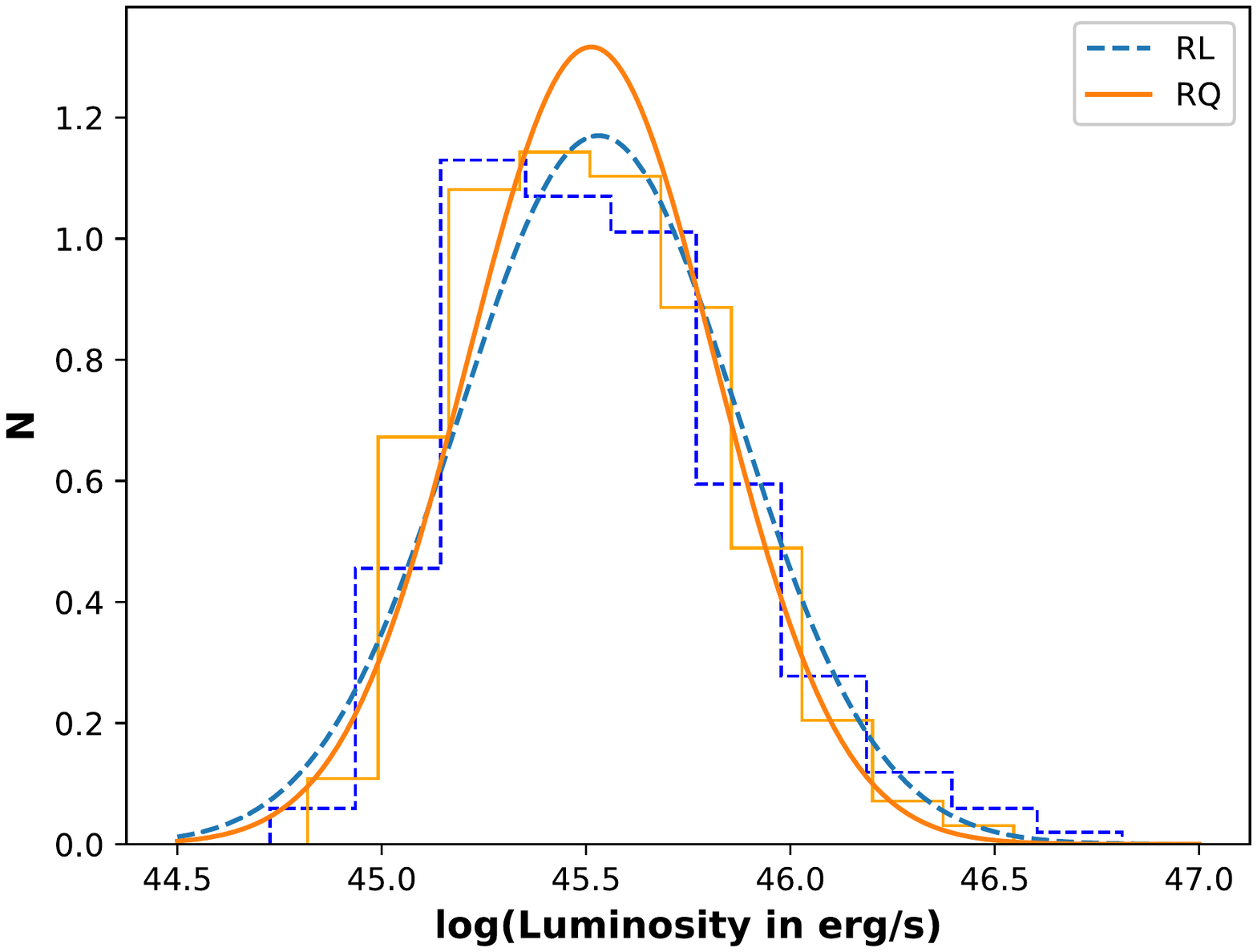}
\vspace{-0.75in}
\caption{Normalised distribution of bolometric luminosity of RLQs and RQQs for the HBL with Gaussian fits. Here we have taken the luminosity in log scale and its unit is $10^{-7}$ watt or erg/s and normalised distribution of bolometric luminosity of RLQs and RQQs for the LBL with Gaussian fits. Here we have taken the luminosity in log scale and it's unit is $10^{-7}$ watt or erg/s respectively.}
\label{Fig.2}
\end{figure}
\section{Implications}
\underline{\it Radio-loud fraction:}  Figure 1 shows the variation of RLF across FWHM. From this figure we can see RLF is increasing with FWHM which implies that in the HBL region quasars are more radio loud.
\\ \underline{\it Analysis of H$\beta$ line properties:} Figure 2 shows normalised distributions of bolometric luminosity of RLQs and RQQs with Gaussian fits for the HBL and LBL respectively.
Now it is clear from luminosity analysis that for the HBL, RLQs and RQQs distributions are different and RLQs are peaking at higher values and for the LBL, RLQ and RQQ distributions are consistent almost.
\\ \underline{\it Orientation of the quasars:} Now to compare our result with other literature we looked for the ratio of 5 GHz core to extended component flux density R as a function of FWHM for the broad H$\beta$ line for quasars plot of Wills $\&$ Browne (1986) From the plot we are getting high logR value for low FWHM. Wills $\&$ Browne (1986) also said that core dominated quasars will have lower FWHM.
\begin{figure}
\begin{center}
\includegraphics[width=2.5in]{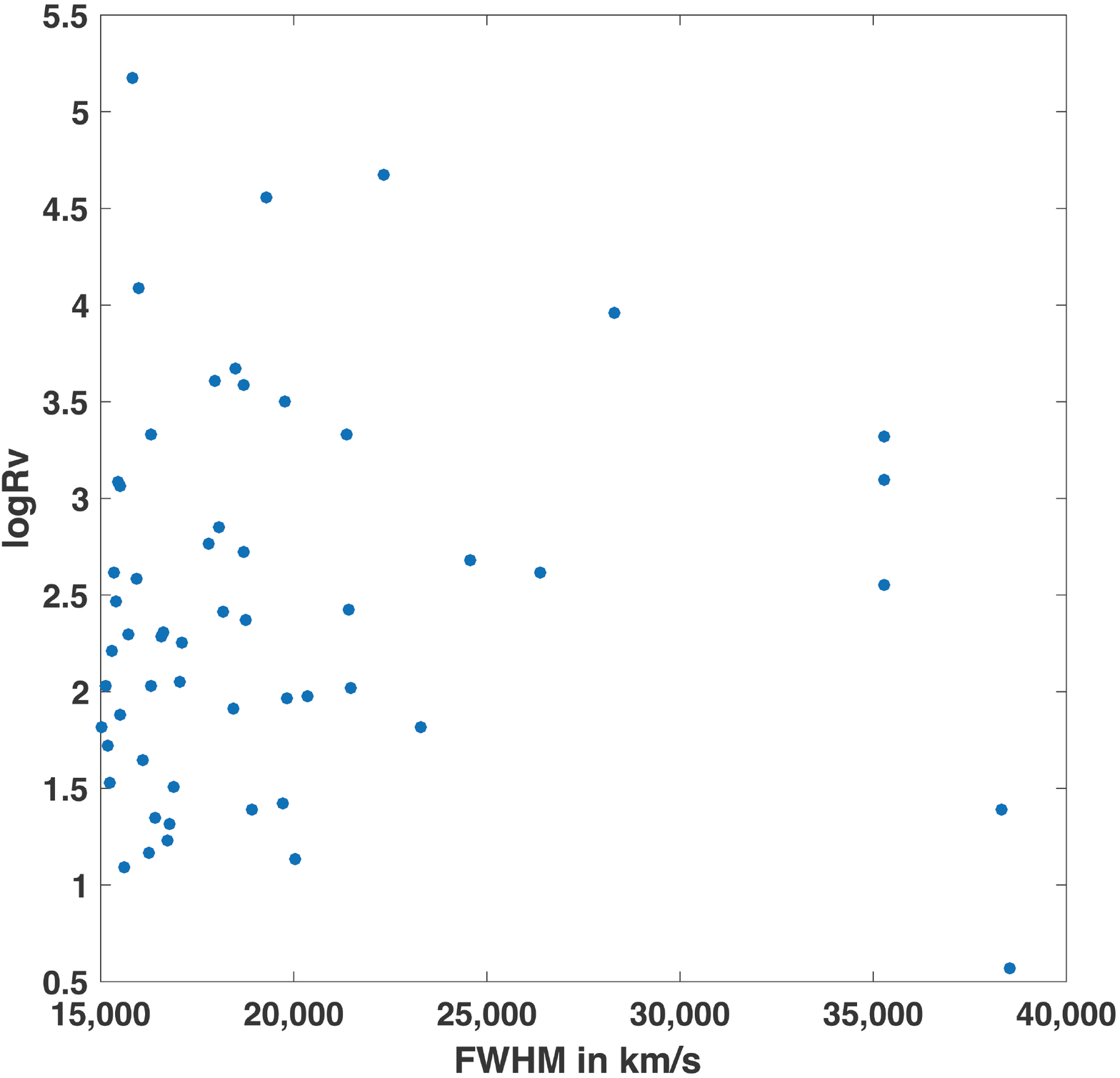}
\vspace{-0.5in}
\caption{log$R_{\mathrm{v}}$ vs FWHM of H$\beta$ line plot for RLQs with FWHM $>$ 15,000km/s.}
\label{Fig.3}
\end{center}
\end{figure}
Then we replot this with our sample but our sample limit is almost beyond their limit. So for the HBL and LBL region, we have calculated log$R_{\mathrm{v}}$. For the HBL (Figure 3) some objects from our sample do not obey the pattern of Wills $\&$ Browne (1986) plot. But the LBL region of our sample is consistent with their plot. 

\section{Discussion}
Our main goal is to investigate whether RLF is consistent across different parameter spaces and here we consider only broad H$\beta$ lines. And we have seen it increases with H$\beta$ FWHM so now to find the reason we chose objects with an exceptionally high full width half maximum (FWHM $>$ 15,000km/s) and compared them with widely used low full width half maximum objects (FWHM $<$ 2,500km/s).\\ 
We compared their bolometric luminosity distributions, and we can see for the HBL RLQs have higher luminosities. Now detection probability of higher luminous objects should be high so this could be a possible reason for getting high RLF in high line widths. We checked for the LBL region also but the distributions for RLQs and RQQs are consistent in that region. More analysis is required to say about the exact reason.
We then tried to compare our result with other literature. We took the ratio of 5 GHz core to extended component flux density R as a function of FWHM for the broad H$\beta$ line for quasars plot of Wills $\&$ Browne (1986) and compared it with log$R_{\mathrm{v}}$ as a function of FWHM for our HBL and LBL sample because Wills $\&$ Brotherton (1995) have suggested a relation between logR and log$R_{\mathrm{v}}$. From our log$R_{\mathrm{v}}$ vs FWHM plot, we saw that some objects do not fall in the pattern described by Wills $\&$ Browne (1986). Further investigation is required for these objects.
\section{Acknowledgement}
We would like to thank Dr. Mike Brotherton and Ms. Jaya Maithili from University of Wyoming, Wyoming and Dr. Suchetana Chaterjee from Presidency University, Kolkata and Ms. Miranda Gilbert of Sul Ross State University, Texas for their valuable inputs and DST-SERB for providing financial support through the ECR grant of Dr. Suchetana Chatterjee.

\end{document}